\documentclass{llncs}
\usepackage{hyperref}
\usepackage[none]{hyphenat}
\usepackage{theorem}
\usepackage{csquotes}

\begin{document}

\title{Brief Announcement:\\
Vehicle to Vehicle Authentication\thanks{\textbf{This is a version that appeared as a brief announcement in 17th International Symposium
on Stabilization, Safety, and Security of Distributed Systems (SSS, 2015).} Partially supported by the Rita Altura Trust Chair in Computer Sciences, Lynne and William Frankel Center for Computer Sciences, Israel Science Foundation (grant 428/11), the Israeli Internet Association, and the Ministry of Science and Technology, Infrastructure Research in the Field of Advanced Computing and Cyber Security. Partially supported by fundings from Polish National Science Center (decision number DEC-2013/09/B/ST6/02251).}}
\author{Shlomi Dolev~\inst{1} \and \L ukasz Krzywiecki~\inst{2} \and Nisha Panwar~\inst{1} \and Michael Segal~\inst{3}}
\institute{Department of Computer Science,
Ben-Gurion University of the Negev, Israel.
\email{ \{dolev, panwar\}@cs.bgu.ac.il}.
\and Institute of Mathematics and Computer Science,
Wroc\l aw University of Technology,
Poland.
\email{lukasz.krzywiecki@pwr.wroc.pl}.
\and Department of Communication Systems Engineering,
Ben-Gurion University of the Negev, Israel.
\email{segal@cse.bgu.ac.il}.}

\maketitle
\date{}
\thispagestyle{empty}

\noindent \textbf{Vehicle Authentication.} In recent future, vehicles will establish a spontaneous connection over a wireless radio channel, coordinating actions and information. Vehicles will exchange warning messages over the wireless radio channel through Dedicated Short Range Communication (IEEE 1609) over the Wireless Access in Vehicular Environment (802.11p). Unfortunately, the wireless communication among vehicles is vulnerable to security threats that may lead to very serious safety hazards. Therefore, the warning messages being exchanged must incorporate an authentic factor such that recipient is willing to verify and accept the message in a timely manner.

\smallskip\noindent\textbf{Our Contribution.} (\textit{i}) Coupling fixed and non-fixed vehicle attributes with the public key, (\textit{ii}) Optical out-of-band communication channel, (\textit{iii}) Adaptation with existing authentication protocols, (\textit{iv}) Verification.

\smallskip\noindent\textbf{Previous Work.} Vehicles utilize wireless communication standard, i.e., IEEE 802.11p Wireless Access in Vehicular Environment (WAVE) based on IEEE 1609 Dedicated Short Range Communication (DSRC). Raya and Haubaux proposed a Public Key Infrastructure (PKI) based vehicle security scheme, however, an active adversary may launch an impersonation attack. Moreover, roadside infrastructure is required to provide the most updated Certificate Revocation List (CRL). Our scheme removes the active participation of roadside units as well as the regional authorities.

\smallskip\noindent \textbf{Problem Statement.} Every vehicles public key is signed by the authorities and can be verified by the receiver, still, an impersonation attack among the moving vehicles is possible. Accordingly, the scenario starts when a vehicle $v_1$ tries to securely communicate with $v_2$ and requests for the public key of $v_2$.
Vehicle $v_3$ pretends to be $v_2$ and answers $v_1$ with $v_3$ public key instead of $v_2$. Then $v_3$ concurrently asks $v_2$ for its
public key. Vehicle $v_1$ is fooled to establish a private key with $v_3$ instead of $v_2$, and $v_2$ is fooled to establish a private
key with $v_3$ instead of $v_1$. Vehicle $v_3$ conveys messages from $v_1$ to $v_2$ and back decrypting and re-encrypting with
the appropriate established keys. In this way, $v_3$ can find the appropriate moment to change information and cause
hazardous actions to $v_1$ and $v_2$.

\smallskip \noindent \textbf{System Model.} (\textit{i}) Light Amplification by Stimulated Emission of Radiation (LASER), (\textit{ii}) LIght Detection And Ranging (LIDAR), (\textit{iii}) Autocollimator, (\textit{iv}) Physically Unclonable Function(PUF).

\smallskip\noindent\textbf{Proposed Scheme.} The proposed approaches for the vehicle to vehicle authentication are summarized as below:

\noindent \textit{Basic Scheme~\cite{ascom}} We propose to certify both the public key and out-of-band sense-able static attributes to enable mutual authentication of the communicating vehicles. Vehicle owners are bound to preprocess a certificate (periodically, possibly during the annual inspection procedure) that signs monolithically both a public key and a list of fixed unchangeable attributes (e.g., license number, brand and color) of the vehicle (extending ISO 3779 and 3780 standards). With such a scheme the vehicle can verify (say by using a camera) that the public key belongs to the specific vehicle to which the connection should be established (rather than a public key of a standing by adversary).

\noindent \textit{Intermediate Scheme~\cite{dyna}} We consider the case of multiple malicious vehicles with identical visual static attributes. Apparently, dynamic attributes (e.g., location and direction) can uniquely define a vehicle and can be utilized to resolve the true identity of vehicles. However, unlike static attributes, dynamic attributes cannot be signed by a trusted authority beforehand. We propose an approach to verify the coupling between non-certified dynamic attributes and certified static attributes via an auxiliary laser communication channel.

\noindent \textit{Sophisticated Scheme~\cite{tech}} At last, we propose to use, the optical Physically Unclonable Function (PUF) to ensure that response from the receiving vehicle is spontaneous, rather than an answer forwarded from another vehicle. Vehicles utilize an out-of-band optical communication channel in order to exchange the PUF stimulated optical challenge and corresponding response from the sender and receiver, respectively.

\smallskip\noindent\textbf{Claims.} We provide an extended proof of the proposed scheme using Spi calculus and BAN Logic, respectively. Our proposed approach adapts the security construction of the conventional Transport Layer Security (TLS) protocol and satisfy two crucial security properties, i.e., (\textit{i}) Authentication: No active or passive adversary would be able to intercept the communication between sender and receiver and (\textit{ii}) Secrecy: No active or passive adversary would be able to reveal neither the secret session messages nor the secret key.

\end{document}